\documentclass[sigconf,anonymous=false]{acmart}
\usepackage{amsmath}
\usepackage{bbm}
\usepackage{booktabs}
\usepackage{balance}
\usepackage{multirow}
\usepackage{pgfornament}

\newcommand{\algname}[1] {{\fontfamily{cmtt}\selectfont {#1}}}

\usepackage{caption}

\DeclareMathOperator*{\argmax}{arg\,max}

\usepackage{bbm}
\usepackage{figsize}
\usepackage{todonotes}
\usepackage{changepage}
\newif\ifworkinprogress
\workinprogresstrue

\ifworkinprogress
    \newcommand{\HA}[1]{\textcolor{purple}{[Himan] #1}}
    \newcommand{\MM}[1]{\textcolor{red}{[Masoud] #1}}
  \newcommand{\RB}[1]{\textcolor{blue}{[Robin] #1}}
  \newcommand{\BM}[1]{\textcolor{orange}{[Bamshad] #1}}

\else
  \newcommand{\HA}[1]{}
  \newcommand{\MM}[1]{}
  \newcommand{\RB}[1]{}
  \newcommand{\BM}[1]{}
\fi

\AtBeginDocument{%
  \providecommand\BibTeX{{%
    \normalfont B\kern-0.5em{\scshape i\kern-0.25em b}\kern-0.8em\TeX}}}
\renewcommand\footnotetextcopyrightpermission[1]{}

\copyrightyear{2021}
\acmYear{2021}
\setcopyright{acmlicensed}\acmConference[UMAP '21]{Proceedings of the 29th ACM Conference on User Modeling, Adaptation and Personalization}{June 21--25, 2021}{Utrecht, Netherlands}
\acmBooktitle{Proceedings of the 29th ACM Conference on User Modeling, Adaptation and Personalization (UMAP '21), June 21--25, 2021, Utrecht, Netherlands}
\acmPrice{15.00}
\acmDOI{10.1145/3450613.3456821}
\acmISBN{978-1-4503-8366-0/21/06}

\begin{document}




\title{User-centered Evaluation of Popularity Bias \\ in Recommender Systems}




\newcommand{\ecmadd}[1]{{\textcolor{blue}{#1}}}
\newcommand{\ecmreplace}[2]{\textcolor{red}{\st{#1}}
\textcolor{blue}{#2}}
\newcommand{\ecmdelete}[1]{\textcolor{red}{\st{#1}}}

\newcommand{\ecmcomment}[1]{\footnote{\textcolor{blue}{[EM: #1]}}}

\author{Himan Abdollahpouri}
\affiliation{%
  \institution{Northwestern University}
  \city{Evanston}
  \country{USA}
}
 \email{himan.abdollahpouri@northwestern.edu}

\author{Masoud Mansoury}
\affiliation{%
  \institution{Eindhoven University of Technology}
  \city{Eindhoven}
  \country{Netherlands}
  }
  \email{m.mansoury@tue.nl}

\author{Robin Burke}
\affiliation{%
  \institution{University of Colorado Boulder}
  \city{Boulder}
  \country{USA}
  }
  \email{robin.burke@colorado.edu}

\author{Bamshad Mobasher}
\affiliation{%
  \institution{DePaul University}
  \city{Chicago}
  \country{USA}
  }
  \email{mobasher@cs.depaul.edu}

\author{Edward Malthouse}
\affiliation{%
  \institution{Northwestern University}
  \city{Evanston}
  \country{USA}
  }
  \email{ecm@northwestern.edu}

\renewcommand{\shortauthors}{Abdollahpouri et al.}

\begin{abstract}

Recommendation and ranking systems are known to suffer from popularity bias; the tendency of the algorithm to favor a few popular
items while under-representing the majority of other items. Prior research has examined various approaches for mitigating popularity
bias and enhancing the recommendation of long-tail, less popular, items. The effectiveness of these approaches is often assessed
using different metrics to evaluate the extent to which over-concentration on popular items is reduced. However, not much attention
has been given to the user-centered evaluation of this bias; how different users with different levels of interest towards popular items
are affected by such algorithms. In this paper, we show the limitations of the existing metrics to evaluate popularity bias mitigation
when we want to assess these algorithms from the users’ perspective and we propose a new metric that can address these limitations.
In addition, we present an effective approach that mitigates popularity bias from the user-centered point of view. Finally, we investigate several state-of-the-art approaches proposed in recent years to mitigate popularity bias and evaluate
their performances using the existing metrics and also from the users’ perspective. Our experimental results using two publicly-available datasets show that existing popularity bias mitigation techniques ignore the users' tolerance towards popular items. Our proposed user-centered method can tackle popularity bias effectively for different users while also improving the existing metrics. 
\end{abstract}




%


\maketitle

\section{Introduction}\label{intro}

Historically, recommendation algorithms were developed to maximize the accuracy of the delivered recommendations to the users. However, as other researchers have noted, there are other important characteristics of recommendations that must be considered, including diversity, serendipity,  novelty~\cite{Vargas:2011:RRN:2043932.2043955,ge2010beyond,castells2011novelty}, and fairness \cite{yao2017beyond}. These characteristics can have enormous impacts on the utility of the recommendations.

In this paper, we focus on the problem of \textit{popularity bias} \cite{ciampaglia2018algorithmic}, the tendency of recommender systems to favor a small set of popular items (items that receive a large number of interactions from the users) in their recommendations, even more than their popularity would warrant, and to, unfairly, disfavor items that lie outside of this set, even when these items are preferred by a significant number of interested users~\cite{park2008long,steck2011item,jannach2015recommenders}.

Recommending long-tail items (items with low popularity) is generally
considered to be valuable to the users \cite{anderson2008long,shani2011evaluating}, as these are items that
users are less likely to know about. Brynjolfsson et al. 
showed that 30-40\% of Amazon book sales are represented by titles that would not normally be found in brick-and-mortar stores at the time of their writing \cite{brynjolfsson2006niches}. They pointed out that access to long-tail items is a strong driver for e-commerce growth: the consumer surplus created by providing access to these less-known book titles is estimated at more than seven to ten times the value consumers receive from access to lower prices online. Long-tail items are also important for generating a fuller understanding of users’ preferences. Systems that use active learning to explore each user’s profile will typically need to present more long-tail items because these are the ones that the user is less likely to have rated, and where the user’s preferences are more likely to be diverse \cite{nguyen2014exploring,resnick2013bursting}. Finally, long-tail recommendation can also be understood as a social good. A market that suffers from popularity bias will lack opportunities to discover more obscure products and will be, by definition, dominated by a few large brands or well-known artists \cite{celma2008hits}. Such a market will be more homogeneous and offer fewer opportunities for innovation and creativity. For all the aforementioned reasons, it is crucial to make sure these long-tail, less popular, items are fairly recommended on a recommender system platform by developing algorithms that give more chance to these items. 


The research on popularity bias and long-tail has mainly taken an item-centered perspective. In other words, the focus has been largely on how recommendation algorithms amplify the popularity of already popular items in the recommendations given to the users. This item-centered focus, however, ignores the fact that not every user is equally interested in popular or less popular content and the popularity bias may affect different users differently. For example, Abdollahpouri et al. \cite{himan2019b}  demonstrated that popularity bias can cause unfair treatment of different user groups based on how interested they are in popular movies. In particular, they defined three user groups \textit{Blockbuster-focused}, \textit{Diverse}, and \textit{Niche-focused}, and showed that the impact of popularity bias on niche-focused users is much greater than the other groups. In their work, they defined \textit{impact} as the extent to which different user groups with a varying degree of interest towards popular movies were exposed to movies with different levels of popularity. The same patterns were found later by Kowald et al. \cite{dominik2019unfairness} on the music domain where users with a lesser interest in mainstream songs were impacted more severely by the popularity bias. These findings showed that popularity bias impacts different users differently and hence it is important to take users into account when mitigating this bias.

There have been numerous attempts for developing algorithms that can help to reduce the over-concentration on popular items. These approaches mainly aim at improving three aspects of the recommendations in order to mitigate this bias: 1) making sure less popular items are recommended \cite{yin2012challenging}, 2) increasing the number of unique recommended items, regardless of their popularity, across all users \cite{adomavicius2011improving,karakaya2018effective}, and 3) making sure different items are \emph{fairly} and \emph{evenly} exposed to different users \cite{vargas2014improving,mansoury2020fairmatch}. See section \ref{metrics_eval} for more details on three different metrics that measure each of these aspects. However, none of these aspects take the users' tolerance towards popularity into account.

In this paper, we investigate popularity bias in recommender systems from the users' perspective. In other words, we are interested in assessing the algorithm's impact on users when it comes to the popularity level (aka mainstreamness) of the recommended items. More specifically, we would like to determine the extent to which the popularity level of recommended items match users' past interest in popular items. In other words, our goal is to measure the degree to which item popularity is personalized for each user. We first show that existing metrics for evaluating the performance of popularity bias mitigation algorithms fail to capture this aspect of the recommendations. We then, propose a metric, \textit{User Popularity Deviation (UPD)} (see section \ref{user_centered_eval}), to measure the impact of popularity bias on different users based on their interest in item popularity. In addition, we investigate several state-of-the-art methods for popularity bias mitigation proposed in recent years and see how they tackle this bias for different users with a varying degree of interest in popular items. Finally, for ensuring a fairer treatment of different users in terms of how they are impacted by popularity bias, we present an approach that calibrates the recommendation lists given to different users according to their tolerance towards item popularity. We show that this popularity calibration technique, not only does it mitigate the popularity bias based on the existing (i.e. item-centered) metrics, it also tackles this bias fairer from the users' perspective according to our measure of user-centered impact.

In summary, we make the following contributions:

\begin{itemize}
    \item We show the limitation of the existing, widely used, metrics for popularity bias mitigation when we take a user-centered point of view.
    
    \item We propose a new metric for measuring the impact of popularity bias on different users according to their interest in item popularity.
    
    \item We investigate several state-of-the-art recent approaches for mitigating popularity bias via both item-centered and user-centered evaluation and point out their limitation in bias mitigation from a user-centered point of view.
    
    \item We present a new approach to mitigate popularity bias that tackles this bias mainly from a user-centered point of view. 
    
    \item We conduct extensive experiments on two publicly available datasets and investigate how our proposed metric and approach compare to the existing ones.
    
    \item We show that user-centered popularity bias mitigation not only does it outperform the existing bias mitigation solutions from the users' perspective, it also reduces the popularity bias from an item-centered point of view.
\end{itemize}

\begin{figure*}
\centering
\SetFigLayout{3}{2}
 \subfigure[Impact on items]{\includegraphics[width=5.5in]{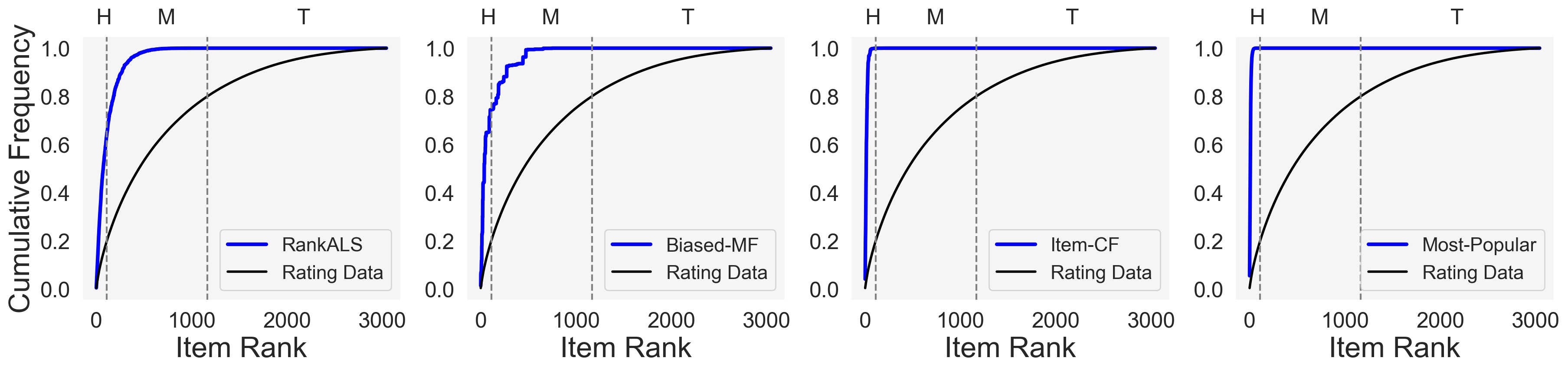}\label{item_groups_chart}}
\subfigure[Impact on users]{\includegraphics[width=5.5in]{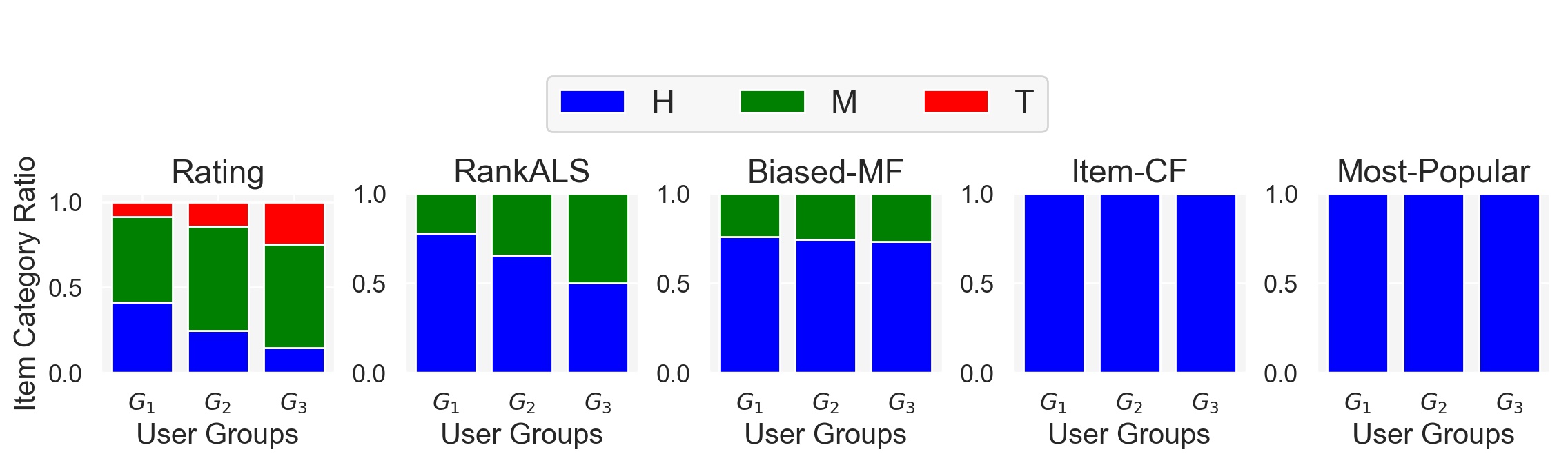}\label{user_groups_chart}}

\caption{The impact of algorithmic popularity bias on items and users. $H$ items are a small number of very popular items that take up around 20\% of the entire ratings. $T$ is the larger number of less popular items that collectively take up roughly 20\% of the ratings, and $M$ includes those items in between that receive around 60\% of the ratings, collectively. Also, $G_1$ are users who are mainly interested in popular items. $G_3$ are users who high interest in niche and less popular items, and $G_2$ are users in between.} \label{base_group_charts}
\end{figure*}

\section{Popularity Bias in Recommender Systems}
Recommendation algorithms are known to suffer from popularity bias where a few popular items dominate the recommendations given to all users. This leads to ignoring the majority of other less popular ones. In this section, we discuss the impact of popularity bias on different items based on their popularity and also  on users based on their tolerance towards item popularity.

\subsection{Impact on Items}
The popularity bias can cause unfair treatment of less popular items. Consider the distributions shown in Figure \ref{item_groups_chart}. These four plots contrast original item popularity in the data and item popularity in the recommendations generated by four well-known recommendation algorithms (\algname{RankALS} \cite{takacs2012alternating}, \textit{Biased Matrix Factorization (\algname{Biased-MF})} \cite{koren2009matrix}, \textit{Item-based Collaborative Filtering (\algname{Item-CF})} \cite{sarwar2001item}, and a simple \algname{Most-Popular} recommendation that recommends most popular items to every user) using the MovieLens 1M \cite{movielens} dataset (see section \ref{data_section} for more details on this dataset). The horizontal axis shows the rank of each item when sorted from most popular to least popular. The black curve shows the cumulative frequency of the ratings for different items at a given rank. As we can see, a few highly ranked items dominate the entire rating history. For instance, only 111 items (less than 3\%) make up more than 20\% of the ratings. In many consumer taste domains, where recommender systems are commonly deployed, the same highly-skewed distribution of user interest is seen. A music catalog might contain artists whose songs have been played millions of times (Beyonc\'{e} or Ed Sheeran) and others whose songs have a much smaller audience (Iranian musician Kayhan Kalhor, for example). These few popular items are referred to as the \textit{Head} (shown by $H$ in the plots) in the literature and make up roughly 20\% of the interactions according to the Pareto Principle \cite{sanders1987pareto}. The rest of the items are usually referred to as \textit{long-tail} and can be divided into two parts \cite{celma2008hits}: \textit{Tail} items ($T$) are the larger number of less popular items that collectively make up roughly 20\% of the interactions, and \textit{Mid} items ($M$) include a relatively large number of items in between that receive around 60\% of the interactions, collectively\footnote{ Of course, one could group items according to their popularity in many different ways but the Pareto Principle, also known as 80-20 rule \cite{dunford2014pareto}, is a well-known concept in Economics for defining wealth (in)equality which is a similar situation if we consider popular items as the richest people and less popular ones as the majority of other people.}. These item groups are shown on the top of each plot partitioning the items based on their popularity into the most popular items ($H$), items with medium popularity ($M$), and less popular items ($T$).

The blue curves in each plot show popularity bias at work across the four algorithms. In the most extreme cases, \algname{Most-Popular} and \algname{Item-CF}, the head of the distribution constituting less than 3\% of the total items make up 100\% of the recommendations given to the users. The other algorithms are only slightly better in this regard, with the \textit{H} items making up more than 64\% and 74\% of the recommendations in \algname{RankALS} and \algname{Biased-MF}, respectively. These plots illustrate the impact of popularity bias on different items and how it creates a ``rich get richer'' dynamic. 

\subsection{Impact on Users}

Although popular items are popular for a reason and many users are interested in them, this interest is not equal for all users. There are many users whose interests might be well outside of the popular items.
As mentioned in section \ref{intro} and shown also by authors in \cite{dominik2019unfairness,himan2019b}, popularity bias could also impact the users of the recommender system by distorting their recommendation lists and not serving them items with the right range of popularity. However, not much attention has been given to this aspect of the popularity bias. To illustrate this impact on users with a varying degree of interest in popular items, we define three equal-size groups of users based on their interest in popular items, similar to how they were defined in \cite{himan2019b} (the ratio of popular items in their profile):

\begin{itemize}
    \item \textbf{Blockbuster-focused Users ($G_1$)}: Users who have a high interest towards popular items. 
    \item \textbf{Diverse Taste Users ($G_2$)}: Users with diverse interest towards popular and less popular items.
    \item \textbf{Niche-focused Users ($G_3$)}: Users who are mainly interested in less popular items.
\end{itemize}
For the rest of the paper, we will use $G_1$, $G_2$, and $G_3$ to refer to the corresponding user groups. 

The first plot in Figure ~\ref{user_groups_chart} shows the ratio of rated items from the three item groups $H$, $M$, and $T$ in the profiles of different user groups in the MovieLens 1M dataset. The vertical axis shows the average proportion of different item groups in the profiles of users in different user groups. Note that all user groups, even $G_1$, have rated many items from the \textit{Mid} (green) and \textit{Tail} (red) parts of the distribution, and this makes sense: there are only a few really popular movies, and even the most blockbuster-focused viewer will eventually run out of them. 

The vertical axis in other plots in Figure~\ref{user_groups_chart} shows the average proportion of different item groups in the recommended items delivered by different algorithms to the users in three user groups. The difference with the original user profiles in rating data is stark, especially in the case of \algname{Most-Popular} and \algname{Item-CF}, where the users' profiles are rich in diverse item groups, the generated recommendations are much less so. \textit{Tail} items do not appear at all. In \algname{Most-Popular} and \algname{Item-CF}, 100\% of the recommendations are from the \textit{Head} category, even for the users with the most niche-oriented profiles (i.e. $G_3$). Generally speaking, none of the user groups (even the ``blockbuster-focused'' $G_1$ users) are getting recommendations that are well-matched to their interests in terms of item popularity, but the ``niche-focused'' $G_3$ users are quite poorly served, getting a steady diet of popular movies in which they are less interested with none of their long-tail ($M$ and $T$ items) interests served. Thus, we see that popularity bias has a differential impact across the user base: all are affected but some much more severely than others. Our measure of the impact of popularity bias should reflect this.

\section{Popularity Bias Mitigation}
To mitigate popularity bias, as mentioned earlier, algorithms often try to improve three aspects of the recommendations:  1) making sure less popular items are recommended, 2) increasing the number of unique recommended items across all users, and 3) making sure different items are \emph{fairly} and \emph{evenly} exposed to different users. Several metrics have been used to evaluate how algorithms improve these aspects of recommendations.

\subsection{Evaluation Metrics} \label{metrics_eval}
We first describe the existing metrics to evaluate the performance of a popularity bias mitigation algorithm. Let $L$ be the combined list of all recommendation lists given to different users (note that an item may appear multiple times in $L$, if it occurs in recommendation lists of more than one user). $L_u$ is the recommended list of items for user $u$. Let $I$ be the set of all items in the catalog and $U$ be the set of all users. 
\subsubsection{Existing metrics: item-centered evaluation}\label{item-centered_eval} The existing metrics for evaluating popularity bias mitigation are mainly item-centered and the most commonly used ones are as follows:
\begin{itemize}
     
     \item \textbf{Average Recommendation Popularity (ARP)}: This measure from \cite{yin2012challenging} and further used in \cite{flairs2019} calculates the average popularity of the recommended items in each list. For any given recommended item in the list, it measures the average number of ratings for those items. More formally:
\begin{equation}\label{eq:arp}
     ARP=\frac{1}{|U|}\sum_{u \in U} \frac{\sum_{i \in L_u}\phi(i)}{|L_u|}
\end{equation}

where $\phi(i)$ is the number of times item $i$ has been rated in the training set (i.e. the popularity of item $i$). Sometimes $ARP$ for an algorithm can be low just because the algorithm has recommended a few extremely non-popular items to everyone. In other words, even if it has not increased the number of unique recommended items, it will have a good (low) $ARP$. For this reason, another metric is also measured to compensate for this drawback of $ARP$ and it is called \textit{Aggregate Diversity} (aka catalog coverage).

     \item \textbf{Aggregate Diversity (Agg\mbox{-}{Div}):} The ratio of unique recommended items across all users:

    \begin{equation}\label{aggdiv}
       Agg\mbox{-}{Div}=\frac{\textbar \bigcup_{u \in U}L_u \textbar}{|I|} 
    \end{equation}

    Higher values for this metric indicate that the recommendations ``cover'' more of the item catalog. This metric is widely used in many prior articles \cite{adomavicius2011improving,karakaya2018effective,antikacioglu2017} to assess the effectiveness of the proposed algorithms to mitigate popularity bias. 
    
    $Agg\mbox{-}{Div}$ is measuring the extent to which an algorithm has increased the number of unique recommended items and hence reduced popularity bias. However, this metric has one limitation: it does not differentiate between whether an item is recommended only once or thousands of times. In other words, as long as an item is included in the recommendations, it will be counted by Equation \ref{aggdiv}. For this reason, another metric is often also looked at to ensure the recommendations are fairly distributed across all the items and that metric is \textit{Gini index}.

    \item \textbf{Gini Index:} Measures the inequality across the frequency distribution of the recommended items. If some items are recommended frequently while other items are ignored, the Gini index will be high. Therefore, lower values for this metric are desirable. 
    
     \begin{equation}\label{gini_ind}
        Gini(L)=1-\frac{1}{|I|-1} \sum_{k=1}^{|I|}(2k-|I|-1)p(i_k|L)
    \end{equation}
    \noindent where $p(i|L)$ is the ratio of occurrence of item $i$ in $L$.

An algorithm that recommends each item the same number of times (uniform distribution) will have a Gini index of 0 and the one with extreme inequality will have a Gini of 1. Gini Index is also used in many prior works \cite{vargas2014improving,eskandanianUMAP2020,mansoury2020fairmatch} to assess the (in)equality of the exposure of different items due to popularity bias.



\end{itemize}

\subsubsection{Proposed metric: user-centered evaluation}\label{user_centered_eval}

\begin{figure}
    \centering
    \includegraphics[height=1.7in]{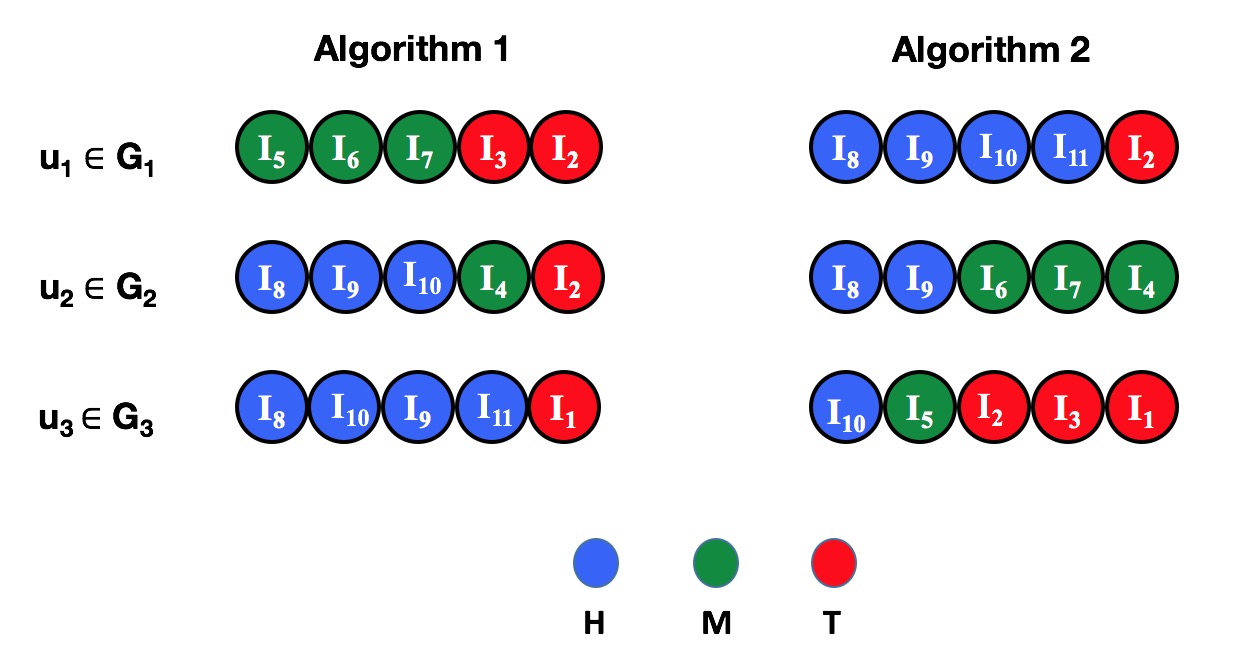}
    \caption{The recommendation lists of size 5 generated by two hypothetical algorithms for three users $u_1$, $u_2$, and $u_3$ belonging to three groups blockbuster-focused ($G_1$), diverse-taste ($G_2$), and niche-focused ($G_3$), respectively. The color of each recommended item indicates the group to which the item belongs.  }
    \label{example}
\end{figure}

All three \textit{ARP}, \textit{Agg-Div}, and \textit{Gini} are assessing popularity bias from the items' perspective. In other words, their focus is largely on measuring how a certain algorithm has treated different items fairly in the recommendations. However, as we mentioned earlier, users are not equally impacted by this bias, and measuring the extent to which different users are affected can be of great importance. Figure \ref{example} shows the recommendations given by two hypothetical algorithms \textit{Algorithm 1} and \textit{Algorithm 2} to three different users $u_1$, $u_2$, and $u_3$ belonging to different user groups $G_1$, $G_2$, and $G_3$, respectively. That means, user $u_1$ is mainly interested in popular items ($H$ items), $u_3$ is mainly interested in non-popular items ($T$ items), and $u_2$ is somewhere in between. Both algorithms have recommended 11 unique items across all three users and the frequency of their appearance is also identical. Therefore, both algorithms have exactly the same $ARP$, \textit{Agg-Div}, and \textit{Gini} and, hence, they will be perceived the same by the existing metrics (item-centered) in terms of popularity bias. However, from the users' perspective, the situation looks quite different. We can see that \textit{Algorithm 1} has not done a good job in matching the recommendations in terms of popularity to the interest of different users: $u_1$ has received many $T$ and $M$ items and $u_3$ has received many $H$ items. \textit{Algorithm 2}, on the other hand, has delivered recommendations to different users that better match the tolerance and tendency of the users in terms of item popularity. This example highlights the importance of having a user-centered evaluation of popularity bias.  

In this paper, we propose a new evaluation metric that captures this algorithmic impact on users in terms of the popularity of the recommended items. We intend to measure the impact of popularity bias on users both overall and on a group-level. 

\textbf{User Popularity Deviation (UPD):} Users have different levels of tolerance towards different item popularity groups. The idea of measuring $UPD$ is to find out how much the recommendations given to each user deviate from that of the user's historical data in terms of interest towards item popularity. For example, if a user has liked 30\% $H$ items, 20\% $M$ items, and 50\% $T$ items, a recommendation algorithm with zero $UPD$ for that user should recommend a list of items that consists of the exact ratio of different item groups. That is, $UPD$ compares the ratio of different item groups in the recommendations given to each user and their corresponding ratios in the historical data.

To do this comparison, we need to compute a discrete probability distribution $P$ for each user $u$, reflecting the popularity of the items found in her profile $\rho_u$ over each item group $c \in C$. For example, the distribution $P$ for the aforementioned example would be $P=\{0.3,0.2,0.5\}$. We also need a corresponding distribution $Q$ over the given recommendation list $L_u$ indicating what item popularity groups are found among the recommended items. For example, if the recommended list contains 70\% items from $H$, 30\% items from $M$, and no item from $T$, then we will have $Q=\{0.7,0.3,0.0\}$. For measuring the interest of each user towards each item popularity group, we use Vargas et al.'s \cite{vargas2013exploiting} measure of category propensity. Specifically, we calculate the propensity of each user $u$ towards each item group $c \in C$ in her profile ($p(c|u)$) and the ratio of such item group in her recommendation list ($q(c|u)$) as follows:

\begin{equation*}
p(c|u)=\frac{\sum_{i \in \rho_u}r(u,i)\mathbbm{1}(i \in c)}{\sum_{c^\prime \in C}\sum_{i \in \rho_u }r(u,i)\mathbbm{1}(i \in c^\prime)} 
\end{equation*}
\begin{equation}
q(c|u)=\frac{\sum_{i \in L_u}\mathbbm{1}(i \in c)}{\sum_{c^\prime \in C}\sum_{i \in L_u }\mathbbm{1}(i \in c^\prime)}
\end{equation}   
$\mathbbm{1}(.)$ is the indicator function returning zero when its argument is False and 1 otherwise. $r(u,i)$ is the rating that user $u$ has given to item $i$.

For any user group $\textsl{g}$, $UPD(\textsl{g})=\frac{\sum_{u \in \textsl{g}}\mathfrak{J}(P(\rho_u),Q(L_u))}{|g|}$. The average \textit{UPD} across different groups of users is:
\begin{equation}
    UPD=\frac{\sum_{g \in G}UPD(\textsl{g})}{|G|}
\end{equation}
where $|G|$ is the number of user groups and $\mathfrak{J}$ is Jensen–Shannon divergence \cite{lin1991divergence}, which measures statistical distance between two probability distributions. Lower values for \textit{UPD} are desirable as it shows the algorithm has less deviation in terms of matching the recommendations with users' tolerance towards items with different levels of popularity.


\subsection{Algorithmic Solutions}\label{algorithm_solutions}

There are several attempts to tackle popularity bias by either modifying the underlying recommendation algorithm or by doing a post-processing step on top of the output of any existing recommendation algorithm. In the former, the popularity of the items is taken into account in the rating prediction \cite{vargas2014improving,sun2019debiasing,abdollahpouri2017controlling,adamopoulos2014over} so the generated recommendations are less biased towards popular items. In contrast, in the latter, a post-processing step is added on top of the output of any existing recommendation algorithm where a larger output recommendation list is taken as input and is re-ordered to extract a shorter final list with improved long-tail properties. Most of the solutions for tackling popularity bias fall into the latter \cite{adomavicius2011maximizing,adomavicius2011improving,antikacioglu2017,flairs2019}.

In this paper, we investigated several papers on bias mitigation published in some top machine learning and recommender systems conferences and, for our analysis, implemented four of them. These four algorithms have all used the existing metrics that we mentioned in section \ref{item-centered_eval} and hence their reported improvements are also based on those metrics. Therefore, they can be great choices to highlight the limitation we mentioned about exiting metrics and how they ignore user-centered evaluation. In addition, we also present a new, user-centered, approach and compare its performance with the previous techniques. These existing approaches are as follows:

\begin{itemize}
    \item \textbf{ReGularization (\algname{RG})} \cite{abdollahpouri2017controlling}: This is a model-based algorithm to mitigate popularity bias that controls the ratio of popular and less popular items via a regularizer added to the objective function of the \algname{RankALS} \cite{takacs2012alternating} algorithm. The algorithm penalizes lists that contain only one group of items and hence attempting to reduce the concentration on popular items. 
    
    \item \textbf{Discrepancy Minimization (\algname{DM})} \cite{antikacioglu2017}:
    In this method, the goal is to improve the total number of unique recommended items, also referred to as aggregate diversity (see Equation \ref{aggdiv}) of recommendations using minimum-cost network flow method \cite{goldberg1997efficient} to find recommendation sub-graphs that optimize diversity. Authors in this work define a target distribution of item exposure (i.e. the number of times each item should appear in the recommendations) as a constraint for their objective function. The goal is therefore to minimize the discrepancy of the recommendation frequency for each item and the target distribution. 
    
     \item \textbf{FA*IR (\algname{FS}) \cite{zehlike2017}:} This method can make a balance between the representation of two groups of items in recommendations: protected and unprotected. Here, the protected items are long-tail ($M \cup T$) items and the unprotected ones are head ($H$) items. The algorithm creates queues of protected and unprotected items and merges them using normalized scoring such that protected items get more exposure. 
     
     
    \item \textbf{Personalized Long-tail Promotion (\algname{XQ}) \cite{flairs2019}:} In this method, inspired by the xQuAD algorithm for query result diversification \cite{santos2010exploiting}, the objective for a final recommendation list is a balanced ratio of popular and less popular (long-tail) items. We specifically included \algname{XQ} since, similar to our approach, it leverages user propensity towards popular items in its calculations and hence it can be categorized as a personalized long-tail promotion technique. However, its main focus is on a balanced distribution of popular versus non-popular items in the recommendation lists, and user propensity is not considered as a first priority but rather as a tie-breaker. Authors of this technique only defined two item categories: short-head and long-tail, with the short-head being $H$ and long-tail being $M \cup T$.
\end{itemize}

Next, we propose an approach that aims at mitigating the popularity bias from a user-centered point of view. 

\subsubsection{Calibrated Popularity (\algname{CP})}
\label{sec:calibration}
Given the proposed metric for a user-centered evaluation of popularity bias mitigation ($UPD$), we intend to propose an approach that addresses the popularity bias from the users' perspective. Our proposed technique, \textit{Calibrated Popularity} (\algname{CP}), is a re-ranking method. We are inspired by the calibrated recommendation algorithm proposed by Steck \cite{steck2018calibrated} that intended to match the distribution of the recommended items over different movie genres to the user's historical interactions. Steck argued that when a user has watched, say, 70 romance movies and 30 action movies, then it is reasonable to expect the personalized list of recommended movies to be comprised of about 70\% romance and 30\% action movies as well. We believe the same argument can be made for the users' preferences towards different item popularity groups. That is, if a user has liked 30 popular items, 20 items with medium popularity, and 50 non-popular items, then it is reasonable to expect the recommendation list to contain 30\% popular items, 20\% items with medium popularity, and 50\% non-popular items. In a nutshell, Calibrated Popularity makes the recommendations more \textit{personalized} in terms of \textit{popularity}.


\algname{CP} algorithm operates on an initial recommendation list $L^{'}_u$ of size $m$ generated by a base recommender to produce a final recommendation list $L_u$ of size $n$ ($m\gg n$) for user $u$. Similar to \cite{steck2018calibrated}, we measure distributional differences in the categories (groups) to which items belong $C=\{c_1$,$c_2$,...,$c_k\}$ between the recommendation list for each user and her profile. For our purposes, these are the three $H$, $M$, and $T$ item groups described above (i.e. $C=\{H,M,T\}$).


Similar to \cite{steck2018calibrated,wasilewski2018intent}, we use a weighted sum of relevance and calibration for creating our re-ranked recommendations. In order to determine the optimal set $L^{*}_u$ from the $m$ recommended items, we use maximum marginal relevance:
\begin{equation}\label{mmr}
    L^{*}_u= \argmax_{L_u, |L_u|= n} (1-\lambda)\cdot Rel(L_u)-\lambda\cdot \mathfrak{J}(P,Q(L_u))
\end{equation}
\noindent where $\lambda$ is the weight controlling the relevance versus the popularity calibration and $Rel(L_u)$ is the sum of the predicted scores for items in $L_u$. Since smaller values for Jensen–shannon divergence, $\mathfrak{J}$, are desirable, we used its negative for our score calculation. As mentioned in section \ref{user_centered_eval}, $P$ and $Q$ are the distribution of popularity of the items found in user's profile and in her recommendation list, respectively.

All five approaches (\algname{RG}, \algname{XQ}, \algname{DM}, \algname{FS}, and \algname{CP}) have a hyperparameter that controls the trade-off between relevance and a second criterion: diversification (incorporating diverse item popularity groups in the recommendations) in \algname{XQ}, fairness (between popular and non-popular items) in \algname{RG} and \algname{FS}, aggregate diversity in \algname{DM}, and user popularity calibration (or deviation) in \algname{CP}. This hyperparameter is shown by $\lambda$ in the result section (section \ref{results_section}).

\section{Experimental Methodology}

\subsection{Data}\label{data_section}
 We have used two publicly-available datasets for our experiments: the first one is a sample of the \textbf{Last.fm} (LFM-1b) dataset \cite{schedl2016lfm} used in \cite{dominik2019unfairness}. The dataset contains user interactions with songs (and the corresponding albums). We used the same methodology in \cite{dominik2019unfairness} to turn the interaction data into rating data using the frequency of the interactions with each item (more interactions with an item will result in a higher rating). In addition, we used albums as the items to reduce the size and sparsity of the item dimension, therefore the recommendation task is to recommend albums to users. We removed users with less than 20 ratings so only consider users for which we have enough data. The resulting dataset contains 274,707 ratings by 2,697 users to 6,006 albums. The second dataset is the \textbf{MovieLens} 1M dataset \cite{movielens}\footnote{Our experiments showed similar results on MovieLens 20M, and so we continue to use MovieLens 1M for efficiency reasons.}. The total number of ratings in the MovieLens 1M data is 1,000,209 given by 6,040 users to 3,706 movies. 
 

 
\subsection{Experimental Settings}

We used 80\% of each dataset as our training set and the other 20\% for the test. As with other re-ranking techniques, our method also needs a base algorithm to generate the initial list of recommendations for post-processing and any standard algorithm can be used. We used \algname{RankALS} and we call it \textit{Base} for the rest of the paper. Similar to \cite{kaya2019comparison}, we set the size of the recommendations generated by the \textit{Base} algorithm to 100 ($m=100$), and the size of the final recommendation list is 10 ($n=10$). We used LibRec \cite{guo2015librec} and \textit{librec-auto} \cite{mansoury2018automating} for generating the recommendations by the \textit{Base} (\algname{RankALS}) algorithm.

\section{Results}\label{results_section}
In this section, we discuss the performance of different algorithms relative to the existing, item-centered, and our proposed, user-centered, evaluation metrics. 

\subsection{Item-centered Analysis}

\begin{figure*}
\centering
\SetFigLayout{3}{2}
 \subfigure[MovieLens]{\includegraphics[width=5.9in]{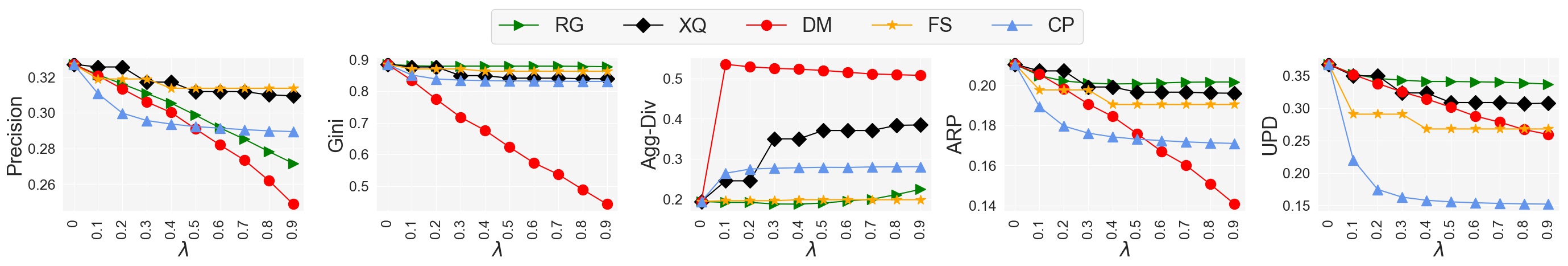}\label{lambda_analysis_ml}}
\subfigure[Last.fm]{\includegraphics[width=5.9in]{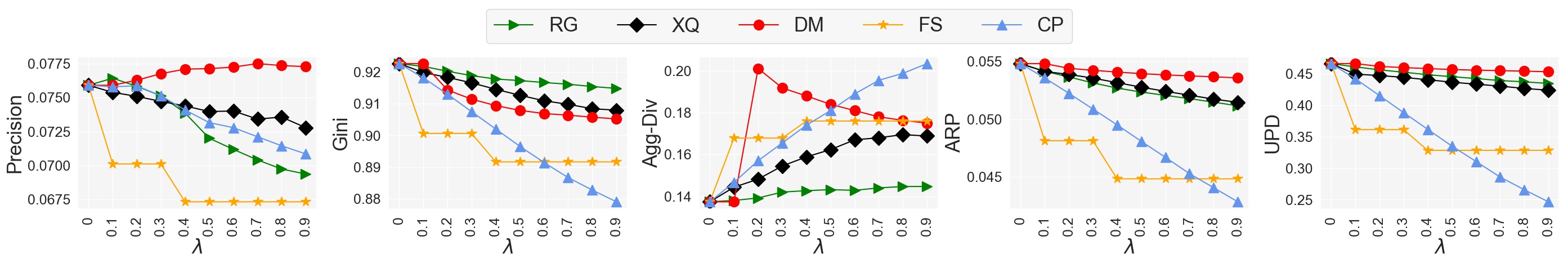}\label{lambda_analysis_lfm}}

\caption{The results of different popularity bias mitigation in terms of Precision, $Gini$ Index, Average Recommendation Popularity ($ARP$), and User Popularity Deviation ($UPD$)} \label{lambda_analysis}
\end{figure*}

Figure \ref{lambda_analysis} shows the performance of five different popularity bias mitigation algorithms, described in section \ref{algorithm_solutions}, when we vary the weight for the diversity of the recommended items in terms of popularity (higher values for $\lambda$ indicates more weight for bias mitigation). 

\begin{figure*}
\centering
\SetFigLayout{3}{2}
 \subfigure[MovieLens]{\includegraphics[width=2.4in]{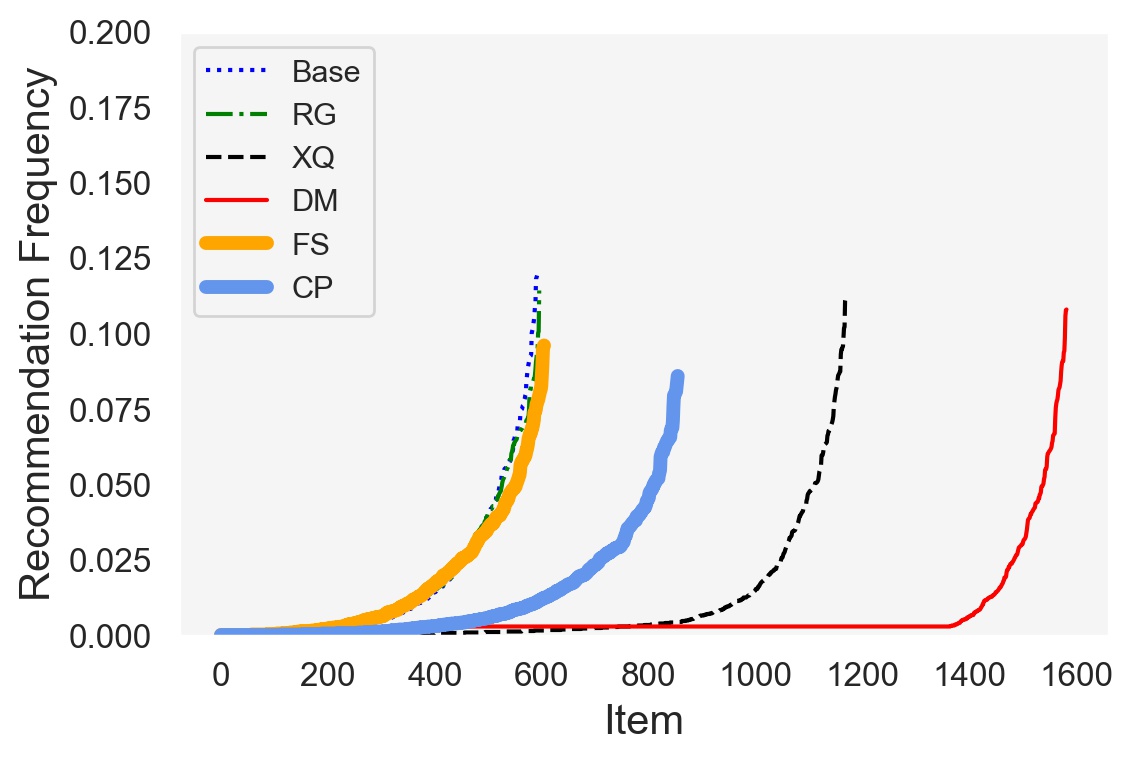}}
\subfigure[Last.fm]{\includegraphics[width=2.4in]{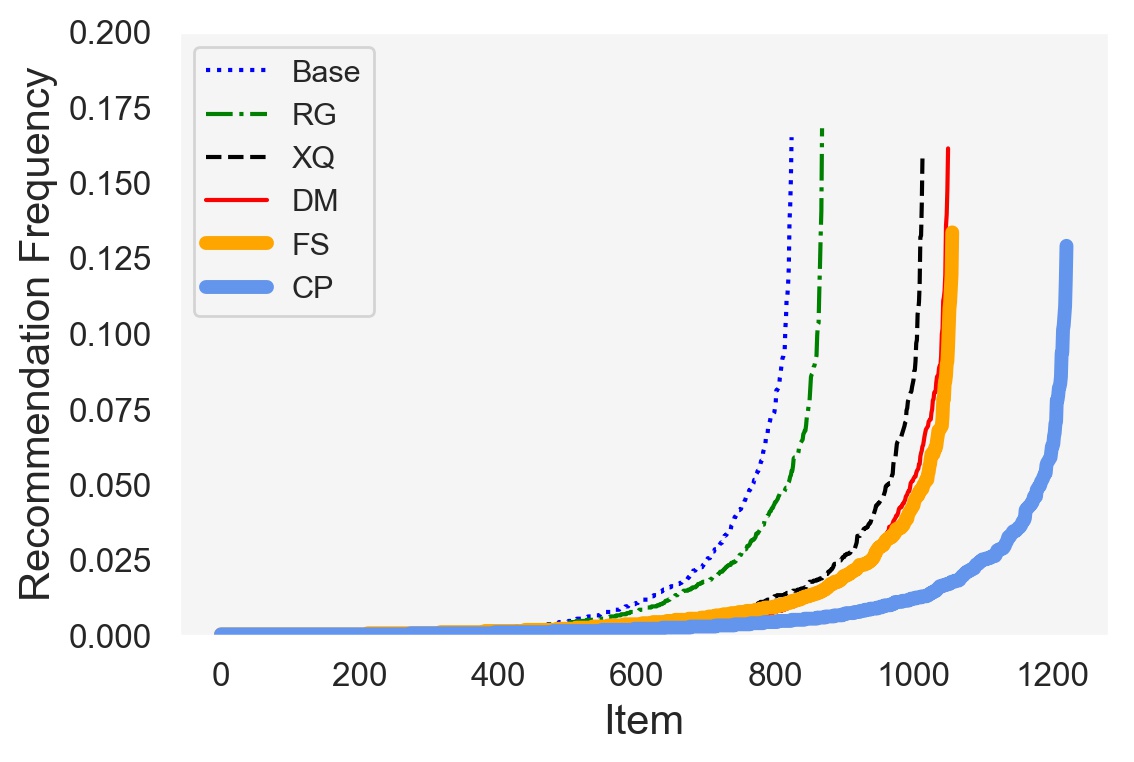}}
\caption{The exposure frequency of different items recommended by different algorithms. The frequencies are sorted for easier comparison.} \label{expoure_sorted}
\end{figure*}

On MovieLens, we can see that all algorithms have lost some degree of precision when we increase $\lambda$, which is consistent with prior work studying a trade-off between relevance and popularity bias mitigation \cite{isufi58accuracy,eskandanianUMAP2020}. Some algorithms such as \algname{XQ} and \algname{FS} seem to perform better in terms of not losing too much precision while other methods such as \algname{DM} and \algname{RG} have lost a more substantial level of precision. \algname{CP} also has lost some precision up to a certain point and it flattens out afterward. In regards to the existing metrics for popularity bias mitigation that we mentioned in section \ref{item-centered_eval}, $ARP$, \textit{Agg-Div}, and $Gini$, we can see that \algname{DM} clearly outperforms all other methods especially for larger values of $\lambda$ (lower $ARP$, higher \textit{Agg-Div}, and lower $Gini$). \algname{XQ} also seems to perform relatively well in terms of \textit{Agg-Div} compared to the other three methods followed by \algname{CP}. However, although both \algname{DM} and \algname{XQ} have performed well in terms of \textit{Agg-Div} and $Gini$, later on in Figure \ref{expoure_sorted} we will discuss how they have achieved that and discuss how these two metrics can be misleading. \algname{XQ} has a better aggregate diversity compared to \algname{CP}, but its $ARP$ is worse. This is primarily due to the fact that \algname{XQ} has recommended more items compared to \algname{CP} but those recommendations are still more concentrated on popular items. So far, in terms of existing metrics ($ARP$, \textit{Agg-Div}, and $Gini$) it seems \algname{DM} would be considered the best as it has achieved high \textit{Agg-Div}, low $Gini$, and low $ARP$. However, as mentioned in section \ref{user_centered_eval} by looking at Figure \ref{example}, these metrics hide the overall picture of how the popularity bias is mitigated from the users' perspective. The $UPD$ results clearly show that \algname{FS} and especially \algname{CP} have performed much better in terms of user-centered popularity bias mitigation. For instance, at $\lambda=0.5$ where both \algname{DM} and \algname{CP} have equal precision and equal $ARP$, the $UPD$ for \algname{CP} is twice as good as the one for \algname{DM} (0.15 versus 0.3) which shows \algname{CP} has matched the recommendations to the users' tolerance towards popularity in a much better way.   

On Last.FM, the results are more promising for \algname{CP} as it has outperformed the other approaches in almost every metric, including the existing item-centered ones and the new user-centered metric. That indicates the fact that it is possible to perform well on the item-centered metrics in terms of popularity bias mitigation when we try to tackle this bias from the users' perspective. It is not completely clear when this would happen and it could be dependent on the characteristics of the datasets as our preliminary analysis showed that the popularity bias in MovieLens is more extreme than that of Last.fm. However, further analysis is needed. Regarding the other existing algorithms, if we ignore a slightly worse precision drop for \algname{FS}, it seems this algorithm is also performing well in the bias mitigation metrics which is consistent with what we saw on MovieLens. In both datasets, it is clear that \algname{RG} does not perform well which confirms the results reported in \cite{abdollahpouri2020popularity} where the author reported that model-based popularity bias mitigation generally does not perform that well compared to the re-ranking ones.

To establish a point of comparison across the algorithms, we picked the results for each algorithm at a particular $\lambda$ that are close to each other in terms of precision, to be able to also analyze the performance of the algorithms in terms of popularity bias mitigation for each user group $G_1$, $G_2$, and $G_3$. We also reported the overall precision, \textit{Agg-Div}, $Gini$, and $UPD$ for each algorithm at this particular $\lambda$ in Table \ref{table} for easier comparison across algorithms. In this table, we can see that on MovieLens, as mentioned earlier, \algname{DM} has performed really well on \textit{Agg-Div} and $Gini$ but not well on $UPD$. More or less is true about \algname{XQ}. The reason can be explained using Figure \ref{expoure_sorted} where we can clearly see how these two metrics can be misleading. This figure shows the exposure each algorithm has given to different items. In other words, the number of times each item is recommended across all users is calculated and plotted. The exposure values are sorted for better and easy-to-interpret visualization. Looking at the plot for Movielens in Figure \ref{expoure_sorted} we can see a flat horizontal line very close to zero point (yet not zero) on the vertical axis for \algname{DM} and, to some extent, for \algname{XQ}. This explains the reason why \algname{DM} and \algname{XQ} achieved better results in terms of \textit{Agg-Div} in Figure \ref{lambda_analysis_ml}; these two algorithms have recommended a larger number of items only a few times since this metric does not care about the frequency of times each item is recommended. The very low $Gini$ for \algname{DM} can be also explained by the same plot; since many items are recommended with the same frequency (equal exposure), Equation \ref{gini_ind} will be misleadingly low. The plot for Last.fm does not show this pattern and it is consistent with the results we saw in Figure \ref{lambda_analysis_lfm} where \algname{DM} and \algname{XQ} did not have a misleadingly good \textit{Agg-Div} and $Gini$. As mentioned earlier, this difference in performance on two datasets could be due to the characteristics of these datasets in terms of existing popularity bias and it needs further research (see section \ref{conclusion}).

\captionsetup[table]{skip=4pt}
\begin{table*}[t]

\centering
\setlength{\tabcolsep}{3pt}
\captionof{table}{The result of different popularity bias mitigation algorithms at one particular $\lambda$ where algorithms have close precision. These values on MovieLens are 0.7, 0.9, 0.6, 0.9, 0.9 for RG, XQ, DM, FS, and CP, respectively. Similarly, on Last.fm these values are 0.9, 0.9, 0.9, 0.6, 0.9. Bold values are statistically significant compared to the second best value in each column with significance level of 0.05 ($\alpha$=0.05). Up arrows indicate larger values are desirable while down arrows indicate smaller values are better.} \label{table}
\begin{tabular}{lrrrrrrrrrrrr}
\toprule
 \multirow{2}{*}{Algorithms} & & \multicolumn{5}{c}{\textbf{MovieLens}} & & \multicolumn{5}{c}{\textbf{Last.MF}} \\\cline{3-7}\cline{9-13}
 
 && Precision$\uparrow$ & Agg-Div$\uparrow$ & $Gini$$\downarrow$ & $ARP$$\downarrow$ & $UPD$$\downarrow$ & & Precision$\uparrow$ & Agg-Div$\uparrow$ & $Gini$$\downarrow$ & $ARP$$\downarrow$ & $UPD$$\downarrow$ \\
 \bottomrule
 
 Base   &&0.327& 0.194   &0.885& 0.21 & 0.368 &&0.076& 0.137  &  0.922 & 0.055 & 0.466\\
RG    &&0.291& 0.20   &0.872 & 0.202 & 0.341 &&0.069& 0.145  & 0.915 & 0.051& 0.435\\
XQ &&0.309& 0.384   &0.839 & 0.196 & 0.308 &&0.073 &0.169  & 0.908 & 0.051 & 0.424\\
DM &&0.300& \textbf{0.519} &\textbf{0.623} & 0.167 & 0.302 &&0.077&0.175  & 0.905 & 0.054 & 0.453\\
FS &&0.313&0.198  &0.863 & 0.190 & 0.268 &&0.067&0.176 & 0.891 & 0.045 & 0.328\\
CP   &&0.289& 0.281  & 0.831 & 0.171 & \textbf{0.152} &&0.071& \textbf{0.203}  &  \textbf{0.879} & \textbf{0.043} & \textbf{0.246}\\
 
\bottomrule
\end{tabular}
\end{table*}

\subsection{User-centered analysis}
\begin{figure*}
\centering
\SetFigLayout{3}{2}
 \subfigure[MovieLens]{\includegraphics[width=5.5in]{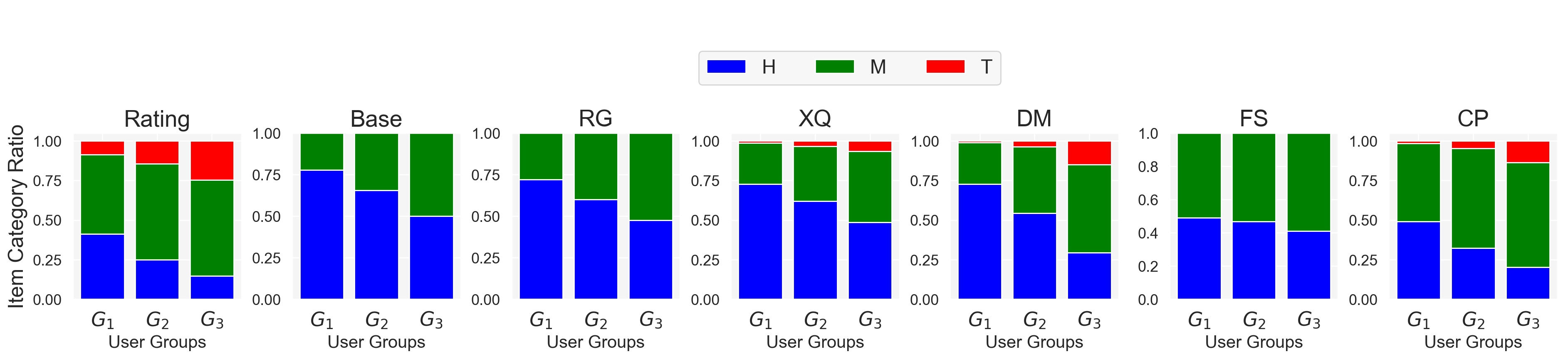}}
\subfigure[Last.fm]{\includegraphics[width=5.5in]{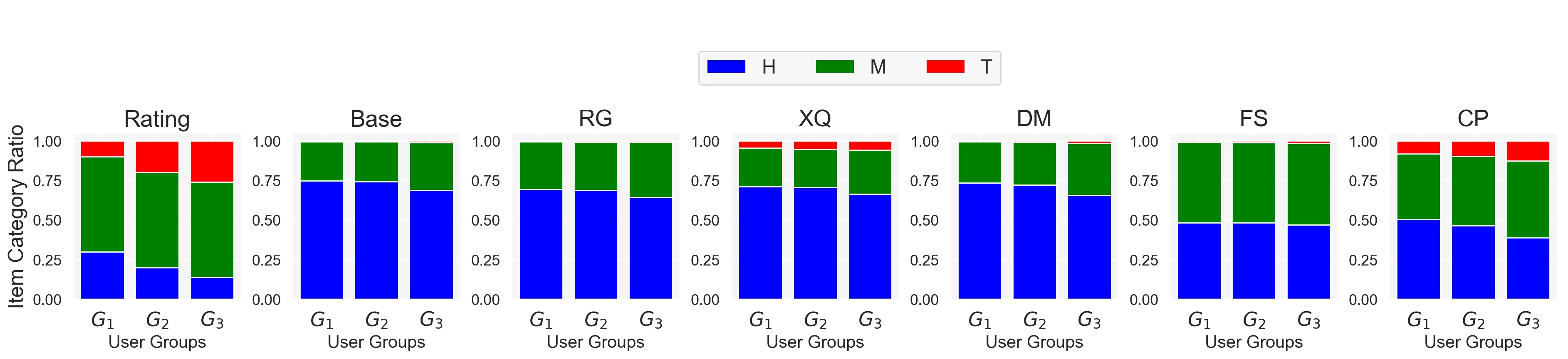}}

\caption{The impact of different popularity bias mitigation algorithms (\algname{RG}, \algname{XQ}, \algname{DM}, \algname{FS}, and \algname{CP}) on different user groups} \label{reranking_barcharts}
\end{figure*}

Figure \ref{reranking_barcharts} shows how each of the algorithms has performed in terms of mitigating popularity bias from the perspective of three user groups $G_1$, $G_2$, and $G_3$. These figures are similar to what we saw in Figure \ref{user_groups_chart} where they compared the performance of four standard algorithms. The charts for the rating data and also for the base algorithm (\algname{RankALS}) are also included for easier comparison. 

Looking at charts for all the bias mitigation algorithms, except for \algname{RG}, it is clear that they all have improved the recommendations in terms of a more balanced ratio of different item groups for three user groups. However, it is obvious that some have done a better job in doing so. For example, we can see that \algname{FS} has certainly increased the exposure of $M$ items for different user groups but not much has been done for the $T$ items where the chart for this algorithm is still missing the $T$ items for all user groups even for $G_3$ in which users have a significant interest towards these less popular items. Among \algname{XQ}, \algname{DM}, and \algname{CP}, we can see that \algname{CP} has matched these ratios much better and the recommendations given to different user groups are closer to what the users in these groups had indicated their interests in the rating data. \algname{DM} and \algname{XQ} have also included $T$ items in the recommendations of different user groups, but a closer look reveals that this inclusion has been done not by reducing over-concentration of $H$ (blue) items but rather by taking away from the $M$ (green) items while the \algname{CP} has balanced things smoother. On Last.fm, \algname{XQ} and \algname{CP} seem to have a better composition of different item groups but \algname{XQ} has done so again by not removing bias from the $H$ items but by shifting the recommendations from the $M$ items towards $T$ items. The extent to which each individual user group is affected by each of these algorithms is worth looking at.

\begin{figure*}
\centering
\SetFigLayout{3}{2}
 \subfigure[MovieLens]{\includegraphics[height=1.5in]{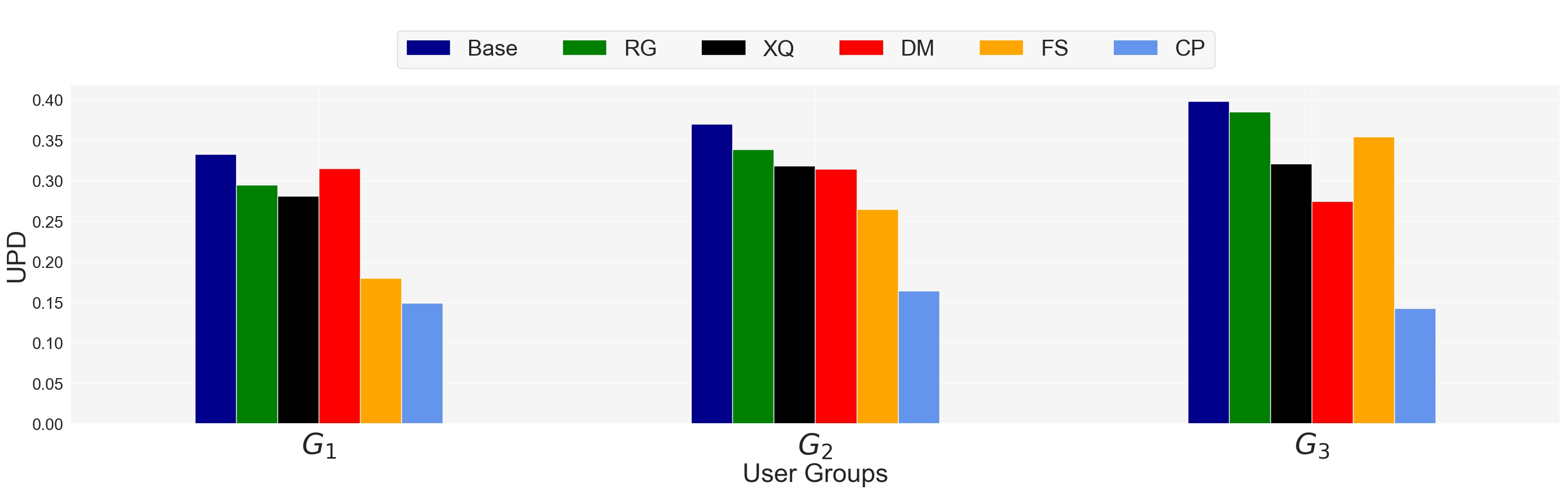}\label{groups_analysis_deviation_ml}}
\subfigure[Last.fm]{\includegraphics[height=1.5in]{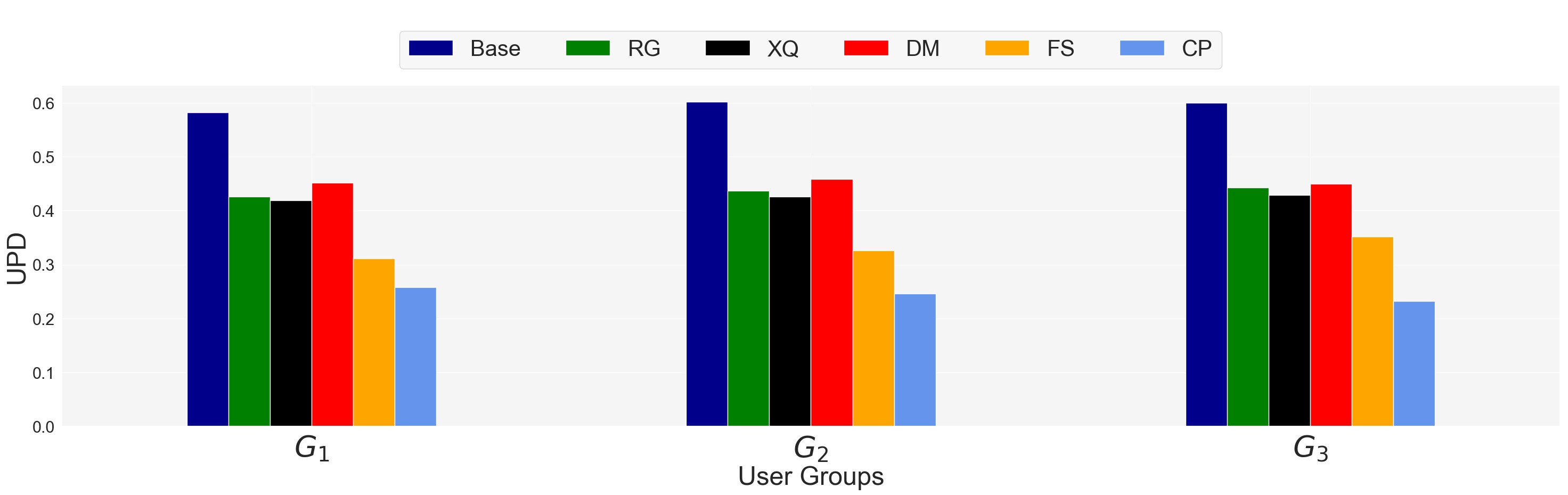}\label{groups_analysis_deviation_lfm}}

\caption{User Popularity Deviation ($UPD$) for different user groups} \label{groups_analysis_deviation}
\end{figure*}

Figure \ref{groups_analysis_deviation} shows the deviation of popularity ($UPD$) experienced by each user group ($UPD(g), g\in \{G_1, G_2, G_3\}$). This plot can further reveal how these different algorithms have performed for each of the user groups. First and foremost, for all user groups, \algname{CP} has the lowest deviation ($UPD$). Another important observation is that different user groups have clearly experienced different $UPD$ values using some algorithms. For example, using \algname{FS}, the deviation for $G_1$ is the lowest followed by $G_2$ and $G_3$ which has the highest $UPD$. That shows that this algorithm has done a better job in terms of removing popularity bias for users who are more interested in popular items ($G_1$) than those who are not ($G_3$). \algname{CP}, on the other hand, has yielded a consistent performance for all user groups giving them persistently low popularity deviation. Another interesting observation that Figure \ref{groups_analysis_deviation} facilitates to perceive is the fact that one algorithm can perform better than another algorithm for one user group but worse for another user group. For instance, on MovieLens, FS has a lower $UPD$ for $G_1$ than \algname{DM} but its $UPD$ for $G_3$ is higher. In other words, if we care about mitigating bias from the blockbuster-focused users' perspective ($G_1$), \algname{FS} outperforms \algname{DM} but if we care more about Niche-oriented users ($G_3$) then \algname{DM} performs better than \algname{FS}.

To gain a more comprehensive picture of how different user groups are affected by popularity bias, looking at the overall $UPD$ across all users may not be enough. For instance, looking at Table \ref{table}, we can see that \algname{XQ} and \algname{DM} have almost similar overall $UPD$ (0.308 vs. 0.302) but, looking at Figure \ref{groups_analysis_deviation_ml}, we can see an interesting pattern: both algorithms have performed almost identical for user group $G_2$ yet \algname{XQ} performs better for users in group $G_1$ and worse for users in $G_3$. 

These are all observations that a typical, item-centered, evaluation procedure of popularity bias mitigation using existing metrics would have not been able to reveal and, hence, we believe, a user-centered evaluation can shed light on many important differences between the algorithms. Nevertheless, the user-centered evaluation proposed in this paper is not to replace the existing item-centered metrics but rather to complement them.

\section{Discussion}
There are two important findings in this paper:

     \textbf{Unfair treatment of different users due to popularity bias.} Consistent with prior works \cite{dominik2019unfairness, himan2019b}, we observed that, in addition to affecting the entire recommendation ecosystem, popularity bias tends to impact different users differently. For instance, we noticed that users with a lesser interest in popular items (niche-oriented users) are more affected by this bias and their recommendations greatly deviate from the range of popularity levels they might expect to receive. This indicates, in addition to an item-centered perspective where the popularity bias is analyzed and mitigated by reducing the bias overall, we also need a user-centered point of view where the impact on different users is investigated and mitigated fairer. 

    \textbf{User-centered vs item-centered evaluation of popularity bias.} In this paper we showed that many of the existing metrics for evaluating popularity bias are mainly item-centered. In other words, they ignore the fact that users have different degrees of interest towards popular items and hence they could be impacted differently by this bias. For instance, we saw that metrics such as aggregate diversity (aka catalog coverage), Gini index, and average recommendation popularity while can be useful in measuring the extent to which an algorithm has reduced popularity bias overall, they fail to capture how the recommendation lists given to different users match their desired range of item popularity. The user-centered metric, $UPD$, we proposed in this paper measures this aspect of the popularity bias. Nevertheless, the user-centered evaluation is not to replace the existing, item-centered, metrics but rather to complement them for a more comprehensive evaluation of popularity bias from both items' and users' perspective. However, the extent to which these two aspects (item-centered vs user-centered) are connected needs further investigation.

    \textbf{\textit{Limitation}: popularity calibration $\neq$ content calibration.} It is worth noting that calibrating the recommendations in terms of popularity does not necessarily guaranty that the recommended items will be matched to the users' overall preferences in terms of content. For instance, a documentary movie and a film noir movie might be both non-popular but a user might be interested in one while disliking the other one. With that said, the popularity calibration method we proposed in this paper does not aim to calibrate the recommendation lists in terms of content but rather only based on the popularity of the items. This popularity calibration is mainly to overcome popularity bias overall and also from the users' perspective to give a more diverse list of recommendations to the user in terms of popularity. Indeed, the popularity calibration can be combined with the genre calibration proposed by \cite{steck2018calibrated} to address this issue. However, this was not the focus of our work in this paper.

\section{Conclusion and Future Work}\label{conclusion}

Popularity bias is a drawback of many recommendation algorithms that favor a few popular items while ignoring the majority of less popular ones. There have been numerous studies to investigate and also mitigate this bias. However, not much attention is given to the impact of this on users and the approaches that can be developed to tackle this bias from the users' perspective. In this paper, we studied the popularity bias from the users' perspective and showed that depending on users' tolerance towards popular items the impact they experience could be significantly different. We also highlighted some limitations of the commonly used metrics for measuring popularity bias mitigation so future usage of these metrics by other researchers can be done with more caution. In addition, we proposed a metric to measure popularity bias from the users' perspective and an approach to mitigate this bias from a user-centered point of view. We compared the performance of our proposed approach with several state-of-the-art approaches for mitigating popularity bias. As expected, the existing approaches, even though have achieved improvements on existing, item-centered, metrics, a further investigation revealed that they did not perform well from the users' perspective. Interestingly, we noticed that our approach which tackles popularity bias from the users' perspective was also able to improve the existing, item-centered, metrics to a great extent.

For future research, we intend to investigate the followings:
\begin{itemize}

    
    \item The \algname{CP} method presented in this paper was relatively simple and mainly for highlighting the points we intended to make about the existing metrics and algorithms for popularity bias mitigation. The user-centered algorithm has, therefore, a great potential for future work and enhancement.
    
    \item We observed that some algorithms performed differently on two different datasets. Our preliminary analysis showed some differences in terms of the degree of popularity bias in the data for these two datasets and this could be one reason for that discrepancy in the performances. One future work could be to investigate how the characteristics of a dataset (e.g. degree of popularity bias, number of users and items, density, or other statistical features) can impact the degree to which a bias mitigation algorithm can tackle popularity bias.
    
    \item Another interesting future work is to study the impact of popularity bias on different user groups over time. Feedback loops are known to amplify popularity bias \cite{mansoury2020feedback}, and the extent to which this amplification differs for different user groups is an interesting question. 
\end{itemize}

\section*{Acknowledgment}
Authors Abdollahpouri and Burke were supported in part by the National Science Foundation under Grant IIS-1911025.

\bibliographystyle{ACM-Reference-Format}
\balance
\bibliography{main}
\end{document}